\documentclass[9pt,twocolumn,twoside]{osajnl}

\journal{optica} 

\setboolean{shortarticle}{true}

\usepackage{lineno}

\title{High-sensitivity quantum sensing with pump-enhanced spontaneous parametric down-conversion}

\author[1,$\dagger$,*]{Chiara Lindner}
\author[1,$\dagger$]{Jachin Kunz}
\author[1]{Simon J. Herr}
\author[1]{Jens Kie\ss ling}
\author[1]{Sebastian Wolf}
\author[1]{Frank K{\"u}hnemann}

\affil[1]{Fraunhofer Institute for Physical Measurement Techniques IPM, Georges-K{\"o}hler-Allee 301, 79110 Freiburg}

\affil[*]{Corresponding author: chiara.lindner@ipm.fraunhofer.de}
\affil[$\dagger$]{These authors contributed equally to this letter.}

\begin{abstract}
Recent years have seen the development of quantum sensing concepts utilizing nonlinear interferometers based on correlated photon pairs generated by spontaneous parametric down-conversion (SPDC). Using SPDC far from frequency degeneracy allows a “division of labor” between the mid-infrared photon for strongest sample interaction and the correlated near-infrared photon for low-noise detection.
The small number of photons provided by SPDC and the resulting inferior signal-to-noise ratio are, however, a limiting factor preventing wide applicability of the novel sensing concept. 
Here, we demonstrate a nonlinear interferometer based on pump-enhanced SPDC with strongly improved emission rates, but maintaining broadband, spontaneous emission. For validation of the concept, we demonstrate high-resolution mid-infrared spectroscopy with near-infrared detection, showcasing the improved accuracy. Although the number of mid-infrared photons is about five orders of magnitude smaller than in classical spectrometers, the sensitivity of the quantum spectrometer becomes comparable, marking an essential step toward real-world applications.
\end{abstract}

\setboolean{displaycopyright}{true}

\begin{document}

\maketitle

\begin{figure}[tb]
	\centering
	\fbox{\includegraphics[scale=0.5]{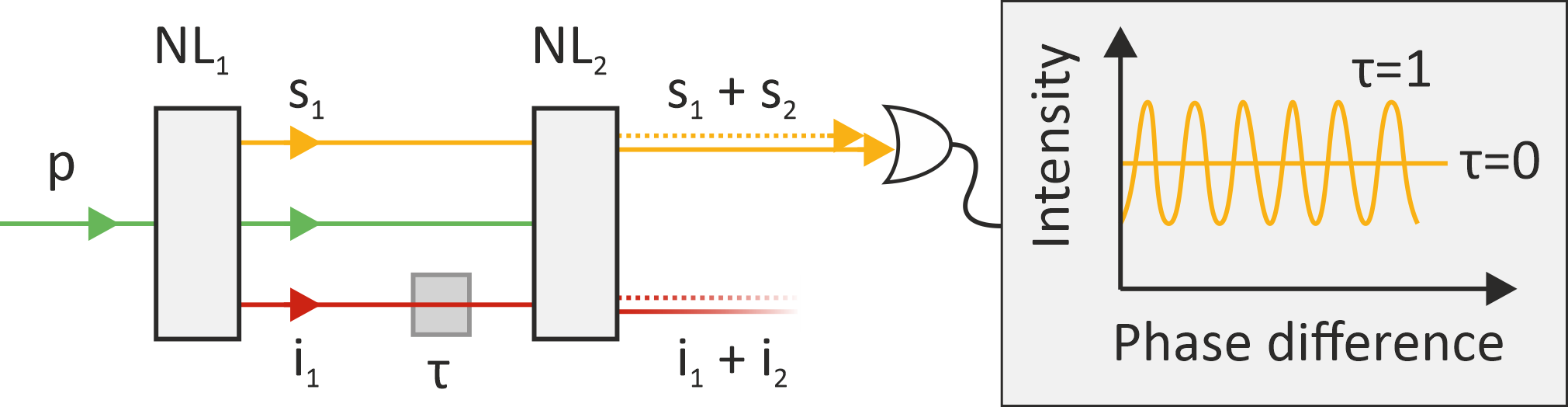}} 
	\caption{Measurement principle of nonlinear interferometers based on the induced coherence effect: A coherent pump beam (green) illuminates two nonlinear crystals (NL$_{1,2}$), which emit correlated signal (orange, s$_{1,2}$) and idler (red, i$_{1,2}$) photons. Due to induced coherence, the interference contrast of the signal light depends on the transmission $\tau$ of the sample inside the idler beam, even though the signal photons have never interacted with it.} 
	\label{fig:principle_sketch}
\end{figure}

The interference effects of correlated photon pairs allow measurements with \textit{undetected photons}~\cite{Lemos.2014}. Hereby, information on the transmission of one photon path can be detected in the interference pattern of their correlated partner photons. 
A common source for correlated photons is spontaneous parametric down-conversion (SPDC), which can be described as a spontaneous decay of pump photons into two photons of lower energy, called signal and idler. These correlated photons can have vastly different frequencies. In a nonlinear interferometer (see Fig.~\ref{fig:principle_sketch}), two of such sources can be superimposed, so that the signal photons (and also the idler photons) from the first and second source become indistinguishable. In this case, both signal and idler photons will show interference modulation due to an effect called \textit{induced coherence}~\cite{Zou.1991}. Due to the low emission rates of SPDC sources, no induced emission takes place. 

Since the photons are correlated, the interference contrast of signal and idler photons is mutually dependent. If the idler photons are absorbed by a sample (with an amplitude transmission $\tau$), the two correlated photon sources become distinguishable and the signal light will show a decreased interference contrast. This way, information of the idler photon path is imprinted on the signal light. Since there is no induced emission of signal or idler light, the interference contrast depends linearly on the transmission of the sample~\cite{Wiseman.2000}. 

Using correlated photon pairs with mid-infrared idler and visible or near-infrared signal wavelengths allows to acquire mid-infrared information with Silicon-based detection of signal light. Silicon-based detectors operate without cooling and offer lower detector noise as well as higher bandwidths than typical mid-infrared detectors. 
This measurement concept has been demonstrated for various applications such as imaging~\cite{Lemos.2014, Kviatkosvky.2020}, optical coherence tomography~\cite{Vanselow.2020} and spectroscopy with \textit{undetected photons}~\cite{Paterova.2018,Lindner.2020,Lindner.2021}.
Nonlinear interferometers are maturing as tools for precise measurements with low light exposure to the sample. However, due to the low number of photons per spectral element emitted by the SPDC process, the signal-to-noise ratio of measurements based on induced coherence is typically limited by shot noise~\cite{Lindner.2021}, which affects their sensitivity.

Several approaches have been pursued for increasing the emission rates of SPDC sources in the context of various applications. 
The brightness of SPDC sources can be enhanced by placing the light source inside an optical cavity, which is resonant for signal or idler light~\cite{JeronimoMoreno.2010}. According to the spectral profile of the resonator, the bandwidth of the correlated photons is drastically decreased, which can be desirable for many applications, such as those which involve coupling to specific transitions of atomic systems~\cite{JeronimoMoreno.2010}. 
Recently, however, non-resonant SPDC sources with large instantaneous signal and idler bandwidths~\cite{Vanselow.2019} have been demonstrated, which, considering sensing applications such as spectroscopy and optical coherence tomography, is an advantage that should be retained.

A concept allowing for both, bright and broadband emission of correlated photons is that of high-gain parametric down-conversion (PDC). It has been demonstrated that by engineering the phase matching condition of the nonlinear medium, the bandwidth and brightness of PDC can be greatly enhanced~\cite{Chekhova.2018}. However, in contrast to nonlinear interferometers based on spontaneously emitting sources, approaches based on PDC receive a nonlinear response of the interference contrast to the sample transmission~\cite{Machado.2020, Wiseman.2000}. In addition, reaching the high-gain regime often requires short-pulse pump lasers, which limits the frequency-anti-correlation of the signal and idler light for applications where spectral information is of interest.
Another option for increased emission rates are waveguide-based SPDC sources, which offer higher efficiencies than macroscopic (bulk) nonlinear crystals due to an improved mode overlap~\cite{Cao.2021}. 
For sensing experiments, however, coupling losses and the narrow mode apperture discourage the multi-pass operation of a waveguide-based nonlinear interferometer with free-space coupling to a sample.

\begin{figure}[t!]
	\centering
	\fbox{\includegraphics[scale=0.5]{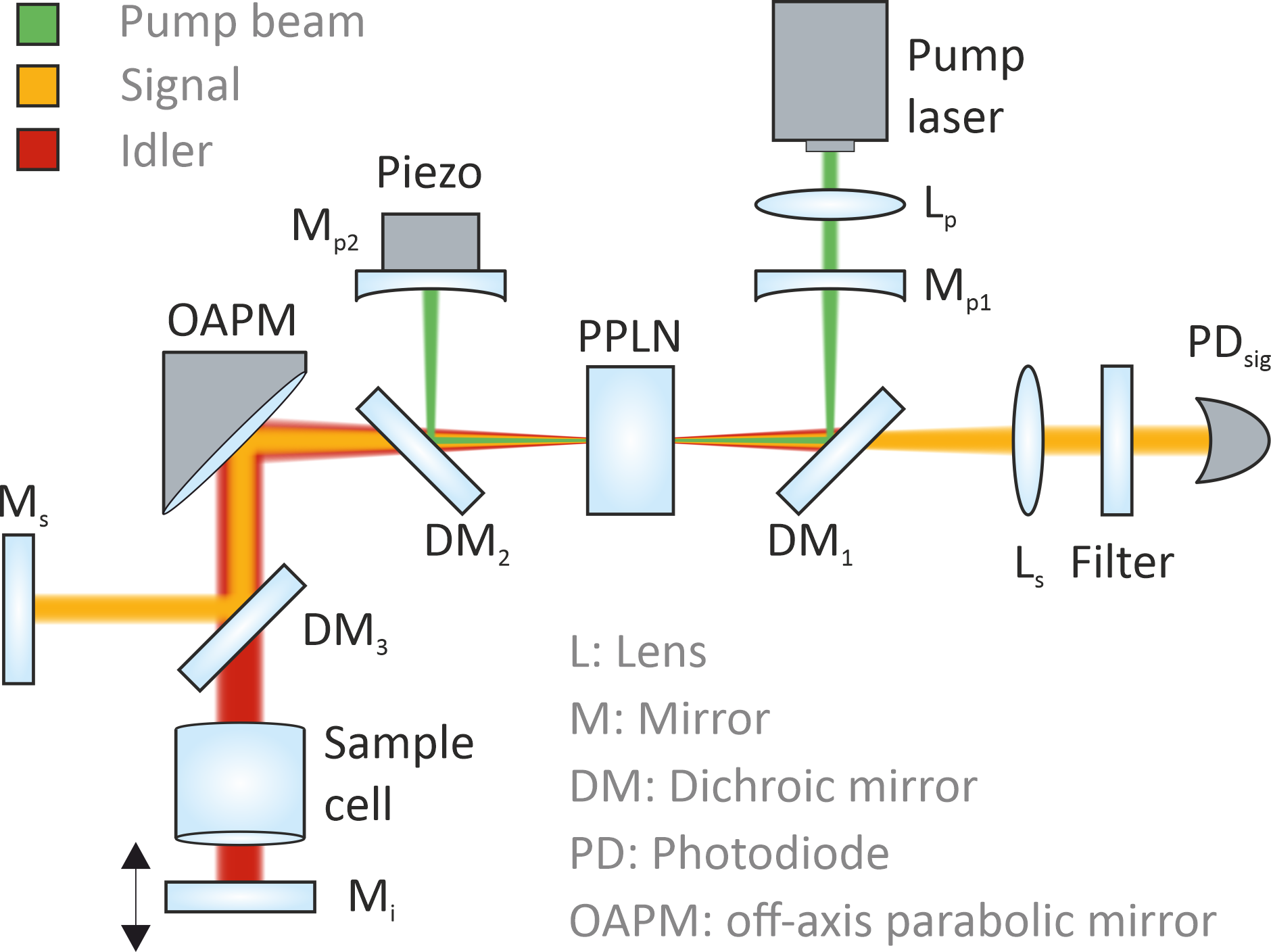}}
	\caption{Schematic setup of the pump-enhanced nonlinear interferometer.}
	\label{fig:setup}
\end{figure}

In a different approach, the SPDC source is placed inside an optical cavity, which is resonant only for the pump light and does not provide any feedback for signal or idler light. This allows to increase the emission rate of SPDC without any modification of the spectral properties of signal and idler light, and without any induced emission. So far, several concepts for pump-enhanced SPDC sources have been investigated: Volz et al. have realized a pump-enhanced SPDC-based source for polarization entangled photon pairs with a low-finesse cavity in semi-monolithic design~\cite{Volz.2001}. Thomas et al. have realized a similar concept with a symmetric cavity~\cite{Thomas.2010}. Katamadze et al. have presented an SPDC source with higher pump enhancement by placing the nonlinear crystal within the cavity of the pump laser~\cite{Katamadze.2013}. This approach, however, complicates possible applications and requires adapting the cavity to the losses of the nonlinear crystal for stable laser operation.

In this letter, we report the first utilization of a pump-enhanced SPDC source in a nonlinear interferometer for sensing with \textit{undetected photons} at enhanced sensitivity.
To this end, the nonlinear crystal is placed within a passive optical cavity resonant for the pump light. With an enhancement factor of about 55, the SPDC emission rate is increased accordingly while retaining the unique features of a nonlinear interferometer based on induced coherence without induced emission. We demonstrate the advantages of this concept for experiments and applications based on sensing with \textit{undetected photons} for the exemplary case of mid-infrared Fourier-transform spectroscopy with near-infrared detection.

\begin{figure}[t!]
	\centering
	\fbox{\includegraphics[scale=0.5]{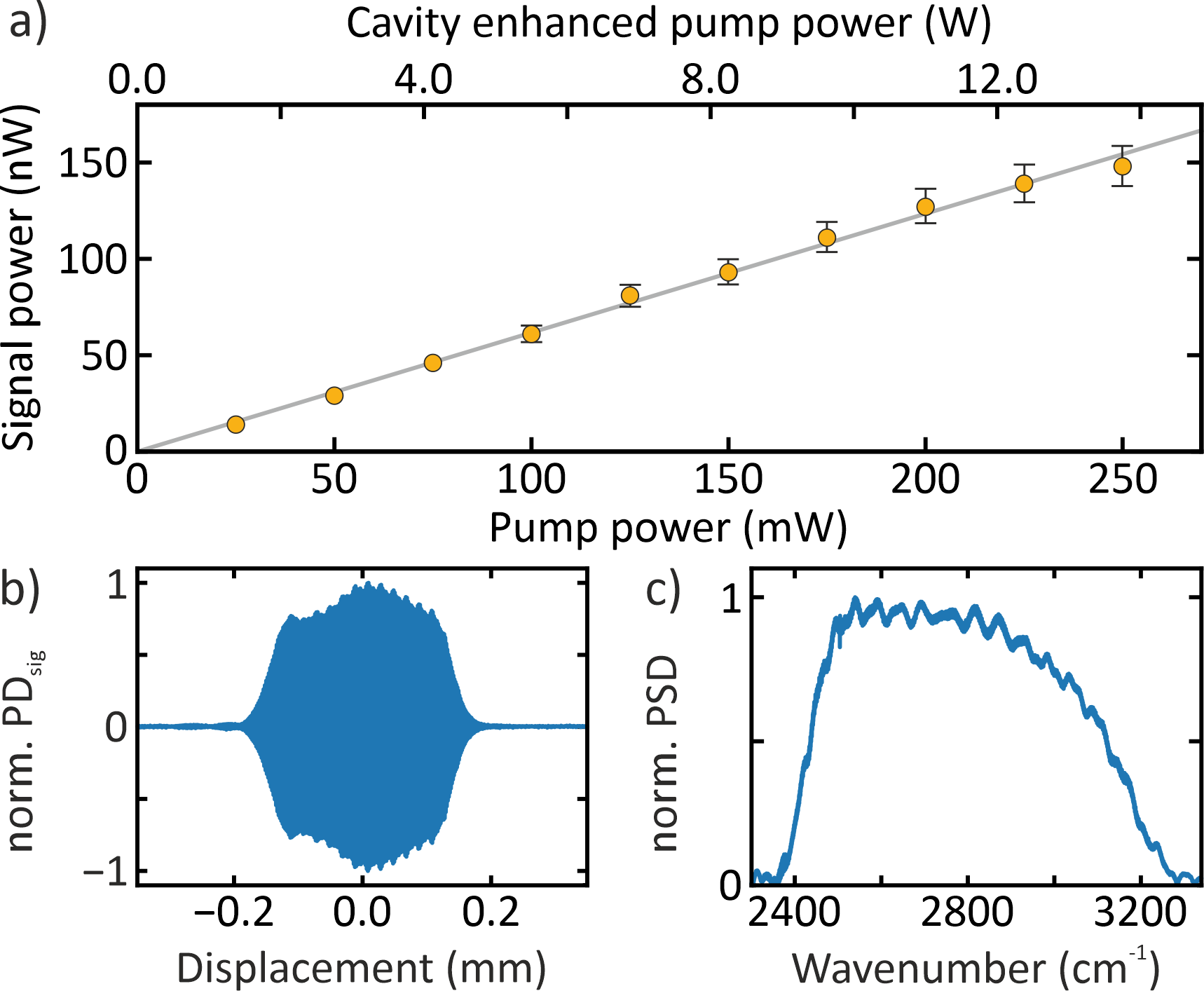}} 
	\caption{Characterization of the pump-enhanced quantum spectrometer:
		a)~Measured SPDC power vs. pump laser power. Top axis shows the calculated cavity enhanced pump power, for an enhancement factor of 55. b)~Detailed view of the measured interferogram (signal light intensity vs. idler mirror displacement). c)~Power spectral density of the reference measurement (sample cell filled with nitrogen).}
	\label{fig:chara}
\end{figure}
In our experiment (Fig.~\ref{fig:setup}) we use a single-frequency pump laser with 775\,nm emission wavelength and up to 250\,mW of output power. The pump laser is coupled into a cavity (formed by the concave end mirrors M$_\text{p1}$ and M$_\text{p2}$) which is piezo-tuned and locked to the laser. The dichroic intra-cavity mirrors (DM$_1$, DM$_2$) are highly reflective for the pump light only, and anti-reflection coated for the involved signal and idler wavelengths. 
The resonator has a finesse of about 290, which results in an enhancement factor of about 55 (taking coupling losses into account).

\begin{figure*}[t!]
	\centering
	\fbox{\includegraphics[scale=0.5]{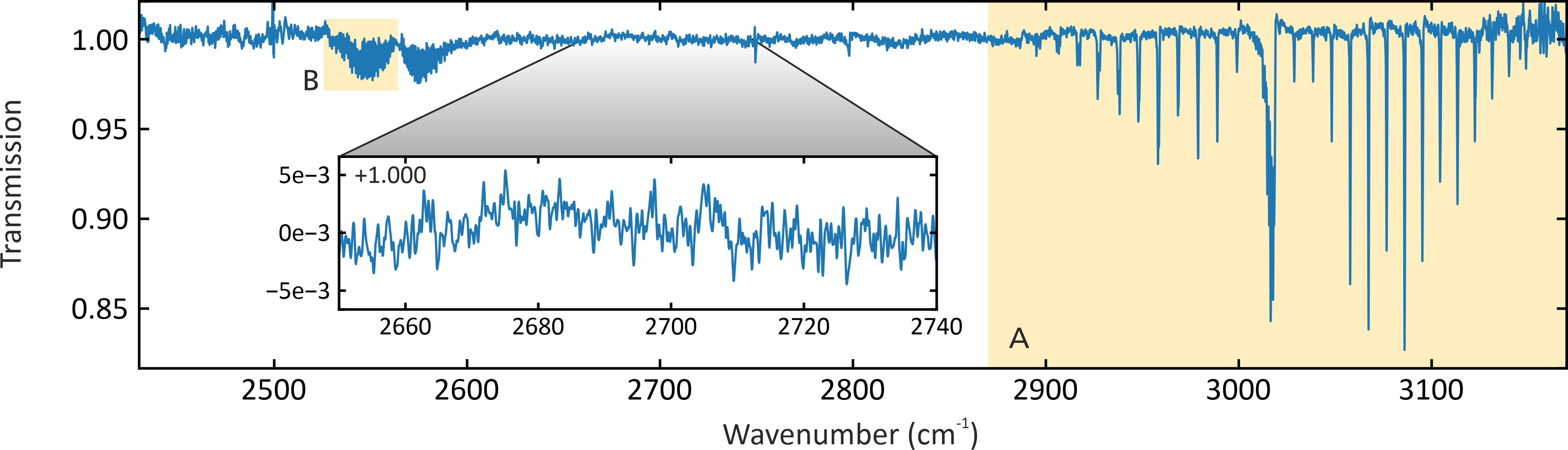}} 
	\caption{Transmission spectrum measured with the pump-enhanced quantum Fourier-transform spectrometer (sample: 0.9\,\% N$_2$O and 0.3\,\% CH$_4$ in nitrogen). The inset shows a detailed view of the 100\,\% line, which allows quantifying the signal-to-noise ratio. The highlighted spectral bands of methane (A) and nitrous oxide (B) are analyzed in Fig.~\ref{fig:zoom}.}
	\label{fig:transmission}
\end{figure*}

As an SPDC source, a 4-mm-long periodically poled lithium niobate crystal (PPLN) is placed at the center of the cavity. The quasi phase matching configuration is designed for near-infrared signal (around 1\,$\text{\textmu}$m wavelength) and mid-infrared idler light emission (around 3.6\,$\text{\textmu}$m wavelength).
The power of the near-infrared signal light emitted by the SPDC source as a function of the pump laser power is shown in  Fig.~\ref{fig:chara}a). The linear dependence confirms that despite the high cavity-enhanced pump power, the SPDC source still emits in the spontaneous, low-gain regime without induced emission.

The SPDC emission to the left-hand side passes the dichroic mirror DM$_2$ and is collimated using an off-axis parabolic mirror (OAPM). The signal light is reflected on a third dichroic mirror (DM$_3$), while the idler light is transmitted and passes through the transmission cell, which can be flushed with sample gas. Both signal and idler beams are reflected on plane mirrors (M$_\text{s}$ and M$_\text{i}$) and imaged back to the SPDC source. Due to induced coherence, the superposition of reflected and direct emission to the right-hand side of the PPLN crystal shows interference modulation when the phase difference of the interferometer is varied by moving mirror M$_\text{i}$. Additional technical details on the nonlinear interferometer setup are given in Supplement 1.

\begin{figure*}[t!]
	\centering
	\fbox{\includegraphics[scale=0.5]{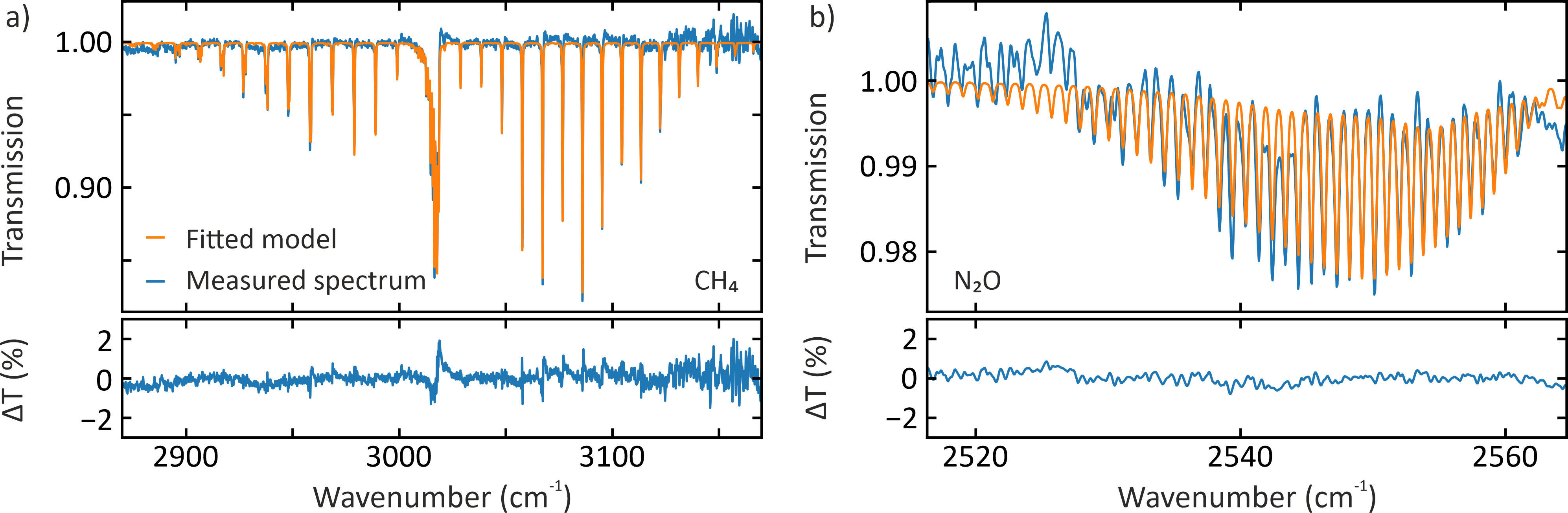}} 
	\caption{Detailed view of the transmission spectrum for an CH$_4$ (left) and N$_2$O (right) absorption band. The measured values are shown in blue, the orange curves show the model function (Eq.~\ref{eq:model}) fitted to determine the concentration of the respective gas. The lower parts show the transmission residuals (difference between measured spectrum and model function).}
	\label{fig:zoom}
\end{figure*}

Recently, it has been demonstrated that the interferometer itself can be used to determine the spectrum of the idler light, in analogy to the measurement principle of a classical Fourier-transform infrared spectrometer (FTIR)~\cite{Lindner.2020,Lindner.2021}. In this measurement setup, the optical path difference between the signal and idler interferometer arm is varied, not only to determine the interference contrast, but to sample the auto-correlation function of the SPDC source. To this end, the idler mirror M$_\text{i}$ is mounted on a high-precision linear stage with a maximum travel range of $\pm$ 20\,mm. As the optical path difference between the signal and idler interferometer arm is varied, a single-pixel Silicon-based photodiode (PD$_\text{sig}$) detects the interferogram of the signal light, which is shown in Fig.~\ref{fig:chara}b). The interferogram is broadened due to the dispersion of the nonlinear crystal~\cite{Lindner.2021}. A Fourier-transform analysis then reveals the broad mid-infrared spectrum of the SPDC source (shown in Fig.~\ref{fig:chara}c)). As in classical Fourier-transform spectroscopy, the spectral resolution is only limited by the maximum optical path difference between the interferometer arms.

In order to test the performance of the pump-enhanced interferometer, we measured the transmission spectrum of a multi-gas sample. At first, a reference measurement is taken wherein the sample cell is filled with pure nitrogen. For the sample measurement, a mixture of 0.9\,\% N$_2$O and 0.3\,\% CH$_4$ in nitrogen is filled into the cell.
Each interferogram is averaged from 50 measurement scans, which were recorded with an acquisition time of 7.6\,s each (resulting in a total measurement time of 380\,s). For these measurements we chose a reduced pump laser power of 100\,mW in order to demonstrate the possible use of a cost-efficient single-frequency diode laser. The interferograms are apodized (cf. Ref.~\cite{Lindner.2021}) with a Gaussian window function with a width of 3.4\,mm (FWHM), which limits the theoretical spectral resolution to about 0.5\,cm$^{-1}$ (FWHM of the instrument line function).

The transmission spectrum calculated from the quotient of sample and reference spectrum is shown in Fig.~\ref{fig:transmission}. The spectrum spans a bandwidth of more than 700\,cm$^{-1}$. 
The transmission spectrum can be used to quantify the signal-to-noise ratio (SNR) and therefore the advantage of pump enhancement. The inset of Fig.~\ref{fig:transmission} shows a detailed view of a spectral range without absorption features of the sample gases. The variance of the measured values around the 100\,\%-line allows estimating the SNR~(cf. Ref.~\cite{Lindner.2021}) to about 570. In comparison to a measurement without pump enhancement, the SNR is increased in agreement to the pump enhancement factor of about 55. For the shot-noise limited measurement, the SNR increases as the square root of the enhancement factor. This allows more sensitive measurements.

Next to the enhanced SNR, the transmission spectrum can also be used to demonstrate the linearity and accuracy of the quantum sensing experiment by comparison to suitable spectroscopic reference data.
For this, we may closer examine the two absorption bands which are highlighted in Fig.~\ref{fig:transmission}: a strong absorption band of CH$_4$ and a much weaker absorption band of N$_2$O. 
The transmission spectrum can be modeled by:
\begin{equation}\label{eq:model}
	T_\text{th}(\tilde{\nu}) = \text{exp}^{-c\alpha(\tilde{\nu}) L} * f(\tilde{\nu})
\end{equation}
using the concentration $c$, the absorption coefficient $\alpha$ based on the HITRAN database~\cite{Gordon.2017}, the length of the sample cell $L$=2\,cm and a convolution with the instrument line function $f(\tilde{\nu})$, which is determined by the Fourier-transform of the apodization function. Figure~\ref{fig:zoom} shows a detailed view of the CH$_4$ absorption band around 3000\,cm$^{-1}$ (a) and the N$_2$O absorption band around 2560\,cm$^{-1}$ (b); the measured transmission values are shown in blue. The orange curves show the model function, which was fitted to the measured values using the respective concentration as a free parameter. The concentrations determined by the fit result to $c_\text{CH4} = 0.3001(11)\,\%$ and $c_\text{N2O}=0.95(2)\,\%$, which is in excellent agreement to the values set in the experiment, even for the weak absorption band of N$_2$O. The high accuracy of the spectrometer is also visible in the small transmission residuals ($\Delta$T) shown in the lower parts of Fig.~\ref{fig:zoom}a) and b).

The multi-gas sensing experiment demonstrates the high quality and sensitivity of spectroscopic measurements with the pump-enhanced nonlinear interferometer. The high spectral resolution allows clearly resolving the rotational lines of the gaseous samples and is only achievable due to the high degree of frequency-anti-correlation of the signal and idler light (which, in turn is due to the low bandwidth of the cw pump laser). The large bandwidth of the SPDC source allows differentiating multiple gases, which demonstrates the benefit of a cavity which is only resonant for the pump light and therefore does not restrict the bandwidth of the signal/idler light. 

The pump enhancement concept requires a narrow-linewidth pump laser for efficient coupling to the cavity, and an active stabilization. With these investments in instrumentation, the pump enhancement allows achieving a higher signal-to-noise ratio, which results in an improved sensitivity of the transmission measurements. This is a key benefit for quantitative quantum sensing measurements. Considering the high spectral resolution, the SNR per measurement time becomes comparable to that of classical FTIR devices~\cite{Griffiths.2006}, while using a much lower light exposure of only about 60\,nW to the sample. In future realizations, the SPDC rates may be enhanced until a significant level of parametric gain is reached. 

We believe that the concept of pump-enhanced sensing will be beneficial to the multitude of measurement concepts with \textit{undetected photons} based on the induced coherence effect. The results prove that the single-pass pump laser beam of previous experiments can be substituted by an enhanced intra-cavity field while retaining the unique properties of the SPDC source and the nonlinear interferometer. 
Pump enhancement allows achieving higher SPDC emission rates already at readily available diode laser pump power levels while preserving the benefits of quantum sensing experiments based on induced coherence. This is a crucial step towards allowing these novel quantum measurement concepts to outperform classical measurement techniques in a wide field of applications.

\begin{backmatter}
\bmsection{Funding} Fraunhofer-Gesellschaft (Lighthouse project QUILT).


\bmsection{Disclosures} The authors declare no conflicts of interest.

\bmsection{Data availability} Data underlying the results presented in this paper are not publicly available at this time but may be obtained from the authors upon reasonable request.

\bmsection{Supplemental document}
See Supplement 1 for supporting content. 

\end{backmatter}





\bibliography{sample}

\bibliographyfullrefs{sample}

\end{document}